\documentclass[prl,twocolumn,floatfix]{revtex4}
\usepackage{graphicx,amsmath,amssymb,multirow}
\begin{document}
\title{Comparative study of the transcriptional regulatory networks 
of E. coli and yeast:
Structural characteristics leading to marginal dynamic stability}
\author{Deok-Sun Lee}
\altaffiliation[Present address: ]{Department of Physics, University of Notre Dame, Notre Dame, Indiana 46556, USA}
\affiliation{Theoretische Physik, Universit\"{a}t des Saarlandes, 
  66041 Saarbr\"{u}cken, Germany}
\author{Heiko Rieger}
\affiliation{Theoretische Physik, Universit\"{a}t des Saarlandes, 
  66041 Saarbr\"{u}cken, Germany}
\date{\today}
\begin{abstract}
Dynamical properties of the transcriptional regulatory network of {\it
Escherichia coli} and {\it Saccharomyces cerevisiae} are studied
within the framework of random Boolean functions.  The dynamical
response of these networks to a single point mutation is characterized
by the number of mutated elements as a function of time and the
distribution of the relaxation time to a new stationary state, which
turn out to be different in both networks. Comparison with the
behavior of randomized networks reveals relevant structural
characteristics other than the mean connectivity, namely the
organization of circuits and the functional form of the in-degree
distribution. The abundance of single-element circuits in {\it
E. coli} and the power-law in-degree distribution of {\it
S. cerevisiae} shift their dynamics towards marginal stability
overcoming the restrictions imposed by their mean connectivities,
which is argued to be related to the simultaneous presence of
robustness and adaptivity in living organisms.
\end{abstract}
\maketitle

\section{Introduction}
Living organisms depend simultaneously on a stable internal
environment and a capability to adapt to a fluctuating external
environment~\cite{causton01}. Since the biological characteristics of
an organism are determined by the interplay between its gene
repertoire and the regulatory apparatus~\cite{babu04}, robustness and
adaptiveness should be generic features of the molecular
interactions composing the gene regulation machinery.  The
organization of the gene transcriptional regulatory network has been
analyzed for numerous organisms, in particular for the prokaryote {\it
Escherichia coli} ({\it E. coli})
~\cite{thieffry98,dobrin04,shenorr02} and the eukaryote {\it
Saccharomyces cerevisiae} ({\it S. cerevisiae})
~\cite{guelzim02,tilee02,luscombe04}. 

Adaptivity of an organism implies the production of different cell
types with different functions from the same genome. This begins with
a regulated transcription by certain proteins, transcriptional factor
(TF)~\cite{orphanides02}. The identification of the target genes for
each TF allows the construction of a gene transcriptional regulatory
network, where the nodes are the genes or operons that produce TF's or
are regulated by TF's, and the directed edges indicate a regulatory
dependence: A directed edge from node $A$ to node $B$ implies that a
TF encoded by gene $A$ is involved in the regulation if the expression
of gene $B$. The expression level of each gene defines the dynamical
state of the network.  To achieve robustness and adaptiveness at the
same time one expects the regulatory network dynamics to be neither
chaotic nor fully insensitive to perturbations, but marginally
stable. Structural characteristics of the network must support these
dynamical features.

Our study reveals specific topological features in the transcriptional
regulatory network architecture of {\it E. coli} and {\it S. cerevisiae} that
shift the dynamics towards marginal stability. {\it E. coli}'s network has
a very low mean connectivity, the number of edges per node, which would lead 
in random networks to a high stability thus deteriorating adaptiveness. 
But we find that single-element circuits which are anomalously rich  
in {\it E. coli}'s network help mutations triggered by random perturbations 
to persist, favoring an unstable dynamical behavior.  
{\it S. cerevisiae} on the other hand has a
sufficiently high mean connectivity which favors chaotic dynamics in random
networks deteriorating stability. Here we find that {\it S. cerevisiae}'s
network has a broad (algebraic) node degree distribution and
we demonstrate the stabilizing effect of this feature upon the dynamics.

Practically, the information about the transcriptional regulatory
network structure - which TF binds to which gene - is available via
the chromatin-immunoprecipitation microarray experiments
~\cite{tilee02}. The question, whether a specific TF enforces or
inhibits the expression of a specific target gene, has to be studied
separately. However, those individual interactions do not necessarily
occur independently and these regulatory interactions are often
combinatorial~\cite{hwa03} and time-, cell cycle-, or
environment-dependent, limiting the available information on the
complete regulation profile. Generic dynamical features then have to
be extracted using model interactions as suggested by
Kauffman~\cite{kauffman}: One digitizes the continuous expression
level to a Boolean variable, $0$ (inactive) and $1$ (active), and
assumes a random static regulation rule for each gene in the form of a
random Boolean function for each gene determining its state at the
next time step by the current states of its regulators. Here {\it
random} means that the output value of these Boolean functions is $0$
or $1$ with equal probabilities.

Based on considerations of random Boolean networks with a fixed number
of regulators $k$ for every element, Kauffman \cite{kauffman}
hypothesized that distinct stationary states - limit cycles -
correspond to different types of cells. This idea got some support
from the agreement of the scaling behavior of the number of
limit-cycles for $k=2$-random Boolean networks and the number of cell
types with respect to the genome size, but was also
debated~\cite{samuelsson03,klemm05}. Among networks with fixed
in-degree, $k=2$ is a critical point distinguishing two different
dynamical phases: stable and unstable against perturbations,
suggesting that the regulatory network dynamics of living organisms is
``on the edge" between order and chaos~\cite{kauffman}.

However, real regulatory networks do not have a fixed in-degree but a
heterogeneous connectivity, even their average in-degree $\langle
k\rangle$ is usually different from $2$. Nevertheless the Boolean
model itself is useful, and recently the effects of the nature of the
regulating rules on the dynamical stability were studied within its
framework~\cite{harris02,kauffman0304}. We propose that the network
structure itself is also relevant for the stability/instability aspect
mentioned before. Therefore we construct a network from the data for
the transcriptional regulatory interactions for {\it E. coli} and {\it
S. cerevisiae}, and study how a point mutation, i.e., an altered
dynamical state of a single element, spreads over the whole network by
inducing another mutation through regulatory interactions.

\begin{figure}
\includegraphics[width=0.8\columnwidth]{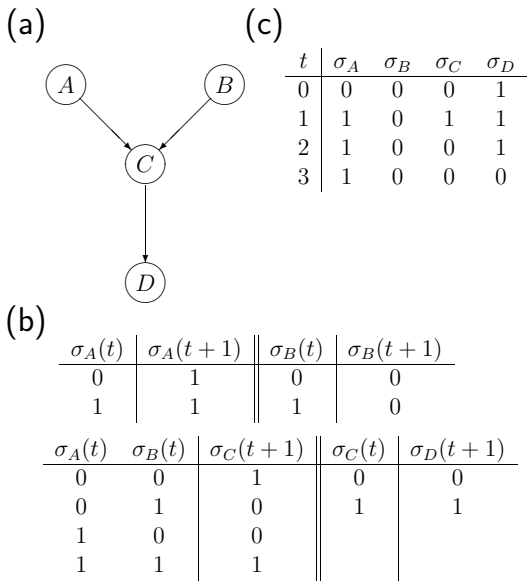}
\caption{An example of Boolean dynamics. (a) A Boolean network of four nodes and three 
  directed edges. Each node has a Boolean variable $\sigma_i$ ($i=A,B,C,D$) 
  (b) Regulating rules $f_i$'s determining the node $i$'s state at time $t+1$ with  
  its regulators' states at time $t$ as input. 
  The nodes $A$ and $B$ have no regulator  
  and their Boolean variables take constant values, respectively, at time $t+1$ 
  regardless of their values at time $t$. 
  (c) An example of the time evolution of those Boolean 
  variables under the regulating rules in (b).}
\label{fig:model}
\end{figure}

\section{Method}
{\it Datasets} --- 
For the transcriptional regulatory network in {\it E. coli}, we used
the data of Ref.~\cite{shenorr02}, which are based on an existing
database, RegulonDB, and enhanced by literature search. The resultant
network consists of $418$ operons and $519$ interactions with $111$
nodes having at least one outward edge. The data for {\it
S. cerevisiae} are taken from Ref.~\cite{tilee02} and were obtained
from the combination of Chromatin Immunoprecipitation and DNA
microarray analysis.  We chose the P value threshold $0.01$, yielding
a network of $4555$ nodes and $12455$ directed edges with $112$ nodes
having at least one outward edge.  Isolated nodes and those possessing
only self-regulation have been excluded in both networks since they
have no interaction with other elements.

{\it Random Boolean functions} --- 
These experimental data establish a directed network $G$ of $N$ nodes,
and we assign a dynamic Boolean variable $\sigma_i$ (that can take on
the values $0$ or $1$ only, corresponding to an inactive or active
state, respectively) to each node $i$.  These dynamical variables
evolve synchronously via $\sigma_i(t+1)=f_i(\sigma_{i_1}(t),
\sigma_{i_2}(t), \ldots, \sigma_{i_{k_i}}(t))$, with the nodes 
$i_1, i_2, \ldots, i_{k_i}$ having the outward edges incident on the
node $i$. The output value of $f_i$ for each input configuration
$\{\sigma_{i_1}(t), \sigma_{i_2}(t), \ldots, \sigma_{i_{k_i}}(t)\}$ 
is $0$ with probability $p$ or $1$ with probability $1-p$, which is
determined at the beginning and not changed with time. If $k_i=0$,
$\sigma_i$ is fixed at $f_i$; $\sigma_i(t+1)=f_i$ regardless of the
value of $\sigma_i(t)$.  The parameter $p$ characterizes the
randomness of the regulating rules: If $p=0$ or $1$, the dynamics is
frozen while the system tends to be disordered with $p=1/2$. An
example network with this Boolean dynamics is given in
Fig.~\ref{fig:model}.

{\it Stability measure} --- 
The stability of a time-trajectory $\Sigma(t)$ is assessed by the
effects of a point mutation $\sigma_i \to 1-\sigma_i$ on the dynamical
evolution of the subsequent states. For this, we choose a 
configuration $\Sigma = \{\sigma_1,\sigma_2,\ldots,\sigma_N\}$, and
prepare its mutant,
$\hat{\Sigma}=\{\hat{\sigma}_1,\hat{\sigma}_2,\ldots,\hat{\sigma}_N\}$,
where $\hat{\sigma}_i = \sigma_i$ for all $i$ except $j$ with $j$
chosen arbitrarily. Evolving $\Sigma$ and $\hat{\Sigma}$ on the same
network with the same regulating rules, we count $n_{\rm m} (t)$, the
number of elements $i$'s with 
$\sigma_i(t)\ne \hat{\sigma}_i(t)$, at
each time step $t$. 
A node with $\Delta \sigma_i(t) \equiv |\sigma_i(t)-\hat{\sigma}_i(t)|>0$ 
is considered as mutated. We average $n_{\rm m}(t)$ over different realizations of 
the regulating rules and different initial pairs of configurations to get the
average, $N_{\rm m}(t)=\langle n_{\rm m} (t)\rangle$, which converges
to its stationary value $N_{\rm m}$. 
For each individual normal-mutant pair $(\Sigma,\hat{\Sigma})$, one can measure 
the relaxation time $t_{\rm r}$ after which $n_{\rm m}(t)$ reaches 
its stationary value. Its distribution $P(t_{\rm r})$ is investigated as well. 

\begin{figure}
\includegraphics[width=\columnwidth]{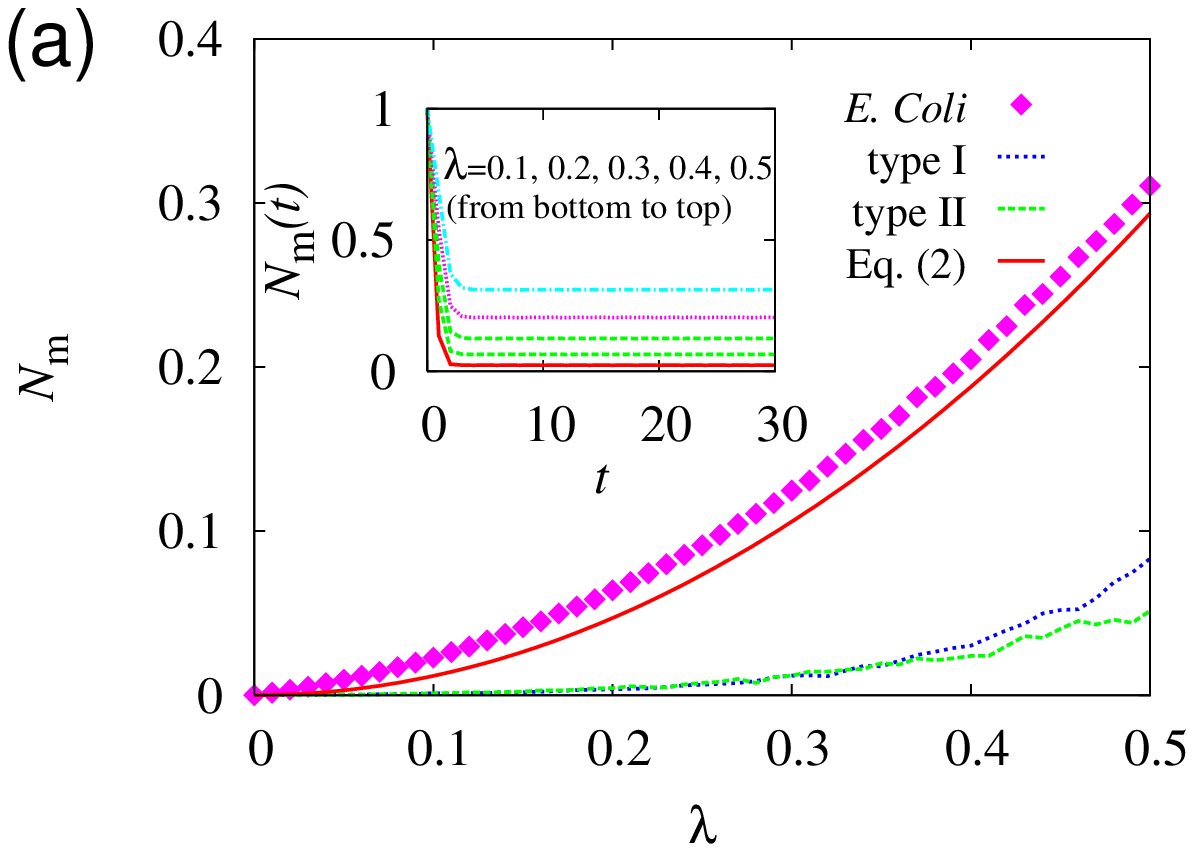}
\includegraphics[width=\columnwidth]{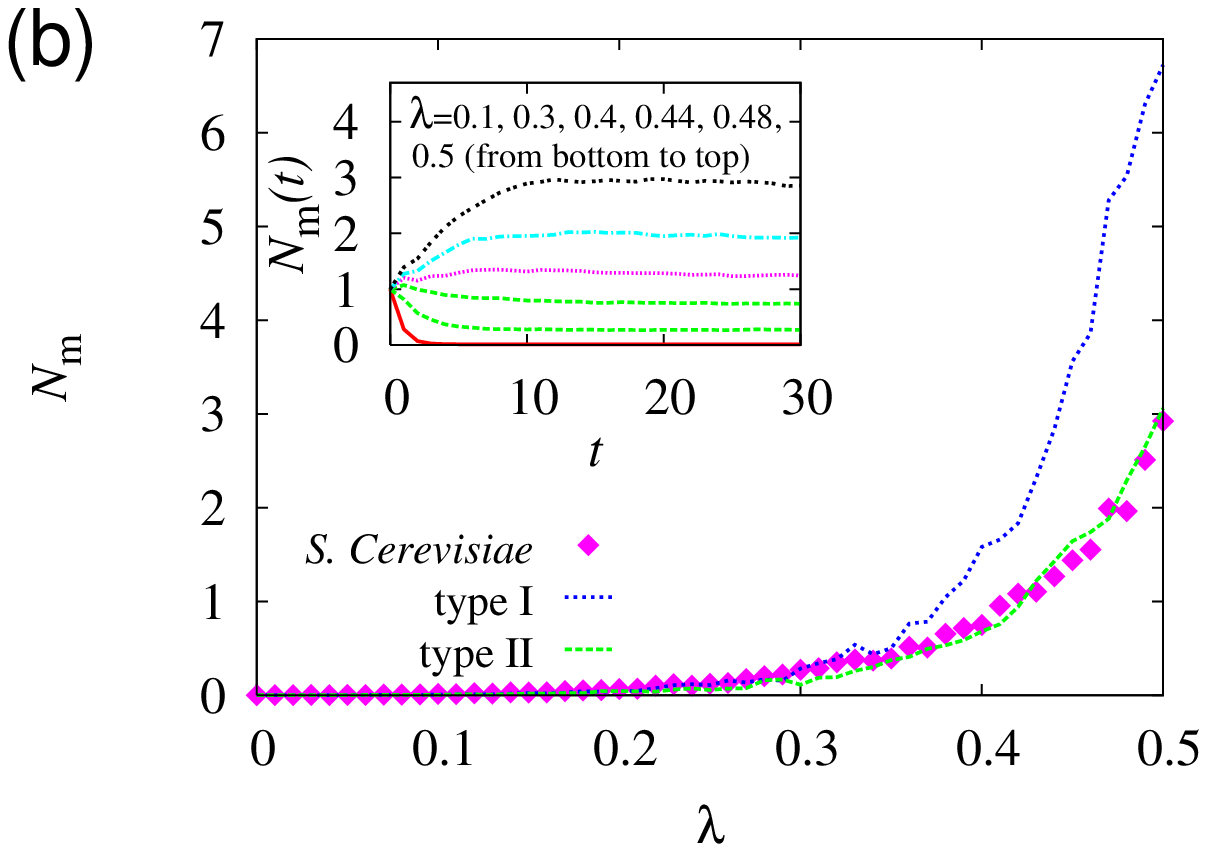}
\includegraphics[width=\columnwidth]{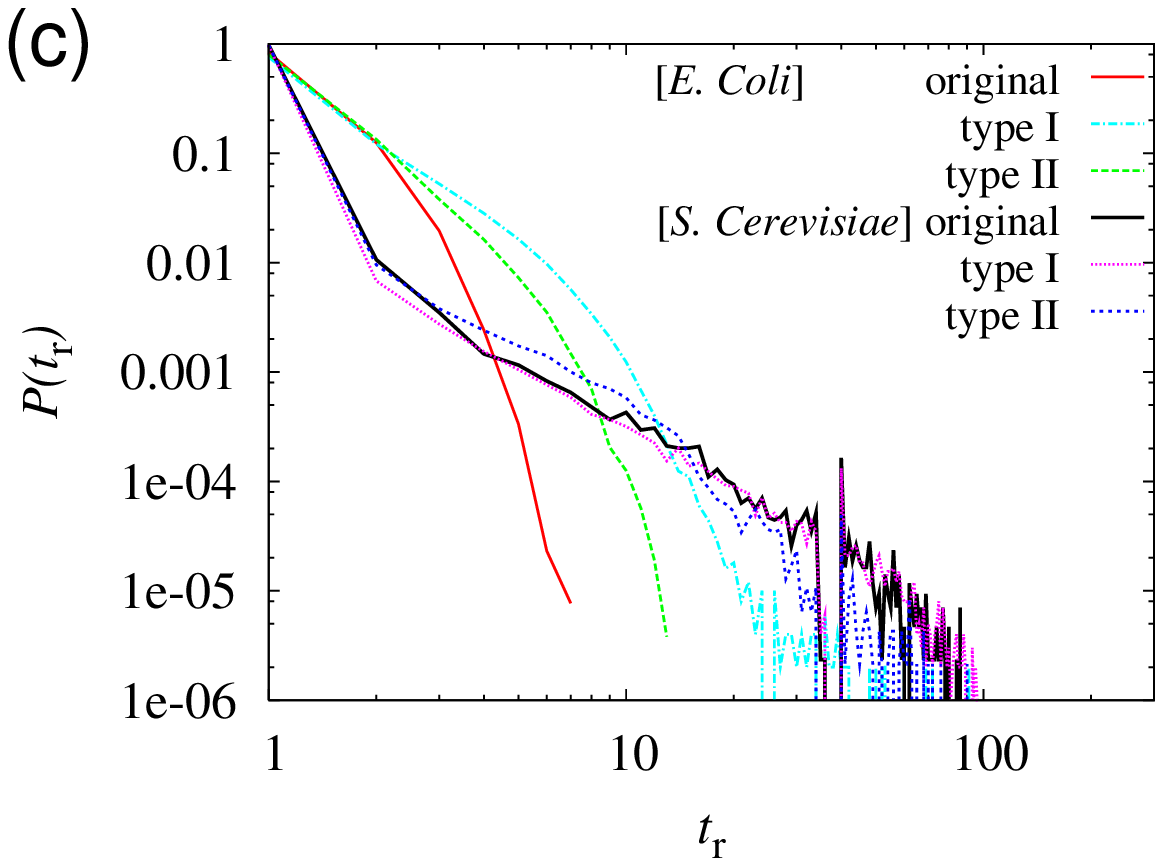}
\caption{Number of mutated elements 
  $N_{\rm m}(t)$ and $N_{\rm m}=\lim_{t\to\infty} N_{\rm m}(t)$ and distribution of the 
    relaxation time $P(t_{\rm r})$. 
  (a) Plot of the stationary value $N_{\rm m}$ versus $\lambda=2p(1-p)$ 
  in the original network and two types of randomized graphs (see the 
  text for the definition) for {\it E. coli}. The data are 
  averages over $10^2$ initial pairs of configurations for each of more than 
  $10^3$ realizations of regulating rules. The approximation given in 
  Eq.~(\ref{eq:ecoliapprox}) is drawn together. The inset shows the time developments 
  $N_{\rm m}(t)$ for selected values of $\lambda$ in the original {\it E. coli} 
  network. (b) The same data as (a) for {\it S. cerevisiae}.
  (c) Plots of $P(t_{\rm r})$ with $p=1/2$ ($\lambda=1/2$) on the original networks and the 
  randomized graphs for {\it E. coli} and {\it S. cerevisiae}.}
\label{fig:NmP}
\end{figure}

\section{Results}
\subsection{Time evolution of the number of mutated elements}
Figure~\ref{fig:NmP} (a) and (b) present
the results for the number of mutated elements 
$N_{\rm m}(t)$ and $N_{\rm m}$.
$N_{\rm m}(t)$ decreases very rapidly 
from $N_{\rm m}(0)=1$ to a much smaller value  for all $p$'s  
in {\it E. coli}. On the other hand, $N_{\rm m}$ for {\it S. cerevisiae} 
increases with time up to a value larger than $1$ for $\lambda \equiv 2p(1-p) 
\gtrsim 0.42$ ($0.3\lesssim p \lesssim 0.7$)  indicating the occurrence of 
a mutation cascade. Both in {\it E. coli} and {\it S. cerevisiae}, 
$N_{\rm m}$ increases with increasing $p$ from $0$ to $1/2$ (or decreasing 
$p$ from $1$ to $1/2$) since the probability that a regulating rule 
yields different output values from different input configurations is 
$2p (1-p)$, which has a maximum at $p=1/2$ and will be denoted by $\lambda$.  
In {\it E. coli}, $N_{\rm m}$ stays smaller than $0.3$,  
indicating that  system-wide mutations are suppressed. 
Figure~\ref{fig:NmP} also shows that in {\it S. cerevisiae} $N_{\rm m}$ is smaller than in 
{\it E. coli} for $\lambda\lesssim 0.2$ but increases with $\lambda$ more rapidly and is 
larger for $\lambda\gtrsim 0.2$.

The functional form of $P(t_{\rm r})$ for $p=1/2$ in Fig.~\ref{fig:NmP} (c) 
is strikingly different between 
{\it E. coli} and {\it S. cerevisiae}: it is exponential for {\it E. coli} and  
a power-law, $P(t_{\rm r})\sim t_{\rm r}^{-1.5(2)}$, for {\it S. cerevisiae}. 
This long tail of $P(t_{\rm r})$ implies that in the case of {\it S. cerevisiae} 
an element can be mutated and recover even at very late times in the dynamics.

\subsection{Mean connectivity}
These differences in the mutation spread dynamics may be
primarily attributed to a difference in the  mean connectivity and 
can be understood by a mean-field approach~\cite{derrida86,aldana03}:
The probability $H(t)=\lim_{N\to\infty} N_{\rm m}(t)/N$ that a randomly chosen node
$i$ is mutated at time $t$, also called the Hamming distance, 
is given in terms of the probability that a  regulator of the node $i$ is mutated,  
which we denote by $\bar{H}(t)$, and the
probability that the regulating rule $f_i$ yields different output values 
from different input configurations, $\lambda$, as 
\begin{eqnarray}
H(t+1)&=& \sum_{k_{\rm in}} \lambda (1 - (1 - \bar{H}(t))^k) P_d(k),
  \nonumber\\
\bar{H}(t+1)&=& \sum_{k,q} \lambda (1 - (1 - \bar{H}(t))^k) \frac{q P_d(k,q)}{\langle q\rangle}.
\label{eq:sc} 
\end{eqnarray}
Here $P_d(k,q)$ is the joint probability that a node has in-degree $k$ and 
out-degree $q$ and is related to the in-degree distribution $P_d(k) = \sum_q 
P_d(k,q)$. $H(t)$ and $\bar{H}(t)$ evolve towards their stationary values 
$H$ and $\bar{H}$. Setting $\bar{H}(t+1)=\bar{H}(t)=\bar{H}$ and expanding 
the second line of Eq.~(\ref{eq:sc}) for small $\bar{H}$, one finds 
$\bar{H}\simeq \bar{H}\lambda \langle kq\rangle/\langle q\rangle  - 
\bar{H}^2\lambda \langle k^2q\rangle/(2\langle q \rangle) + 
\mathcal{O}(\bar{H}^3)$ 
provided $\langle q\rangle$, $\langle kq\rangle $, and $\langle k^2
q\rangle$ are all finite. Therefore $\bar{H}$ and $H$ are zero for
$\lambda$ smaller than a critical value $\lambda_c$ with
$\lambda_c=1/K$ and $K\equiv \langle kq\rangle/\langle q\rangle$ and
non-zero otherwise. The expression $\lambda_c=K^{-1}$ for the critical
point holds true as long as $K$ is finite. Since the Hamming distance
$H$ can be positive only if $K>2$, $N_{\rm m}\simeq HN$ for finite $N$
should be small in {\it E. coli} that has the value $K\simeq 1.08$ and
can be large, of order $N$, for $\lambda\gtrsim 0.42$ in {\it
S. cerevisiae} that has $K\simeq 2.35$.
Although the Hamming distance is not necessarily of order $N^{-1}$ 
at $\lambda_c$, one finds the
value of $\lambda$ for which $N_{\rm m}=1$ very close to the value
$K^{-1}\simeq 0.42$ in the latter.  
The in-degree $k$ and the out-degree $q$ show no significant correlation 
in the two networks according to our analysis not presented here, 
that is, $P_d(k,q)\simeq P_d(k)P_d(q)$ , which yields $\langle kq \rangle 
\simeq \langle k\rangle \langle q\rangle$ and $K\simeq \langle k\rangle$. 

\subsection{Comparison with randomized networks}
Next we studied the same dynamics in two kinds of randomized networks
derived from the regulatory networks of {\it E. coli} and {\it
S. cerevisiae}. The first type of randomized graphs (type I) are
constructed by the repetition of removing an edge connecting nodes
$v_1$ and $w_1$ and creating a new one between $v_2$ and $w_2$, where
both $v_1$ and $v_2$ had at least one outward edge and the node pair
$v_2$ and $w_2$ were not connected before this change. Thus these
type-I randomized networks have the same number of nodes, edges, and
TF's as the original networks, but the edges connect randomly-chosen
pairs of TF and target gene. Our results for $N_{\rm m}$ and $P(t_{\rm
r})$ are shown in Fig.~\ref{fig:NmP}. For the type-I randomized graphs
derived from {\it E. coli}, $N_{\rm m}$ is substantially suppressed as
compared with the original network. In the type-I random graphs
derived from {\it S. cerevisiae}, $N_{\rm m}$ increases much more
rapidly passing $\lambda\simeq 0.3$. The relaxation time distribution
for the random graphs from {\it E. coli} is broader than for the
original network but still decays faster than that for {\it
S. cerevisiae}. The type-I randomization does not change significantly
the relaxation time distribution for {\it S. cerevisiae}.

The type-II randomized graphs we considered are constructed by
exchanging the end points of two edges: Two randomly chosen edges $e_1
= (v_1, w_1)$ and $e_2 = (v_2, w_2)$ are replaced by $e_1' = (v_1,
w_2)$ and $e_2' = (v_2, w_1)$, respectively. These graphs preserve the
joint degree distribution $P_d(k,q)$, but their local connectivity
patterns may be different from that in the original network.  We
present the plots of $N_{\rm m}$ and $P(t_{\rm r})$ in
Fig.~\ref{fig:NmP}. This type-II randomization does not change the
relaxation time distribution for {\it S. cerevisiae} neither. Thus
much faster decay of the relaxation time in the original and
randomized networks for {\it E. coli} than in those for {\it
S. cerevisiae} can be ascribed to the much lower mean connectivity,
$\langle k\rangle \simeq 1.24$, of the former than that of the latter,
$\langle k\rangle \simeq 2.73$. Interestingly the quantities $N_{\rm
m}$ and $P(t_{\rm r})$ for these randomized graphs agree well with
those for the original network of {\it S. cerevisiae}, but not for
{\it E. coli}: This implies that it is the degree distribution that is
mainly responsible for the spread of mutation in {\it S. cerevisiae}
while other (local) structural factors must be important in {\it
E. coli}.

\begin{figure}
\includegraphics[width=\columnwidth]{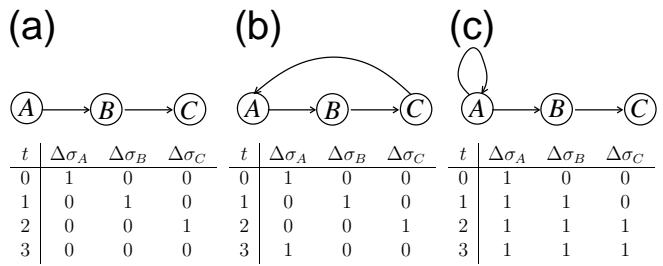}
\caption{Network structure dependence of mutation spread. 
  The regulating rules are given by 
    $f_i(\sigma)= \sigma$ or $1-\sigma$ for nodes $i$'s with one input and 
    $f_i = 1$ or $0$ for nodes $i$'s with no input. Thus a mutated regulator 
    necessarily makes its target node mutated at the next time step. Time evolution of 
    $\Delta \sigma_i = |\sigma_i - \hat{\sigma}_i|$ for each node is shown in 
    tables. 
    (a) No circuit (tree structure).  All nodes recover at $t=3$ and thus the Hamming 
    distance $H$ is zero. (b) A circuit 
of length $3$. The point mutation circulates with period $3$, resulting in $H=1/3$. 
(c) A single-element circuit together with tree structure. All 
nodes are mutated at $t=2$ and thus 
$H=1$.}
\label{fig:tree-circuit}
\end{figure}

\begin{figure}
\includegraphics[width=0.9\columnwidth]{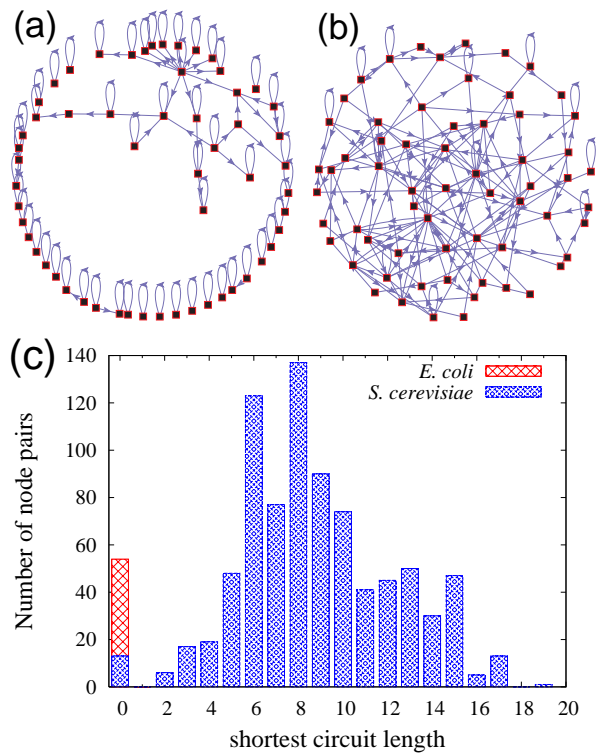}
\caption{Organization of the core in {\it E. coli} and {\it S. cerevisiae}. 
(a) Core of {\it E. coli}. It consists of $57$ nodes and $84$ edges. (b) 
  Core of {\it S. cerevisiae}. It has $63$ nodes and $167$ edges. (c)
Histogram of the shortest circuit lengths. 
In {\it E. coli}, a circuit longer than $1$ 
is not observed but all $54$ circuits are single-element ones. 
In {\it S. cerevisiae}, $836$ pairs of nodes 
among all possible $1953$ pairs in the core are connected by circuits 
and the shortest circuit length ranges from $0$ to $19$.}
\label{fig:core}
\end{figure}

\subsection{Abundance of single-element circuits in {\it E. coli}}
One might expect that circuits (directed closed paths) in the
regulatory network play an important role for the spread of mutations,
because in networks with a tree-structure, i.e., without circuits,
point mutations spread without circulation and a node that is mutated
will recover at the next time step and never become mutated again as
indicated in Fig.~\ref{fig:tree-circuit} (a). The nodes on a circuit,
on the other hand, can return to a mutated state even after recovery
[Fig.~\ref{fig:tree-circuit} (b)]. The nodes lying on circuits or
those on bridges connecting distinct circuits can in principle switch
their status permanently and thus they can be considered as comprising
a core in the dynamics of mutation spread.  As a subnetwork including
all such circuits and the bridges connecting them, we define the core
of a network as the maximal subgraph in which each node has at least
one inward edge coming from and at least one outward edge incident to
an element of the core. 

By deleting the edges having at either end a node that does not meet
the requirement for the core elements, we found the core subnetwork in
the regulatory networks of {\it E. coli} and {\it S. cerevisiae}. Note
that if an edge has the same node at both ends, the node, which
regulates itself, becomes the element of the core. The relevance of
the core to the mutation spread dynamics can be understood e.g., by
investigating the relaxation time distribution $P(t_{\rm r})$ in {\it
S. cerevisiae} depending on the location of the initial point
mutation. Our analysis shows that initial mutations in the core lead
to a qualitatively equal (power-law with the same exponent)
distribution of the relaxation time. On the other hand, initial
mutations in the output module, consisting of all nodes that have
inward edges coming from the nodes in the core and their edges, decay
very fast since the output module has a tree structure and cannot
cause mutations in the core.

The organization of the core turns out to be very different in {\it
E. coli} and {\it S. cerevisiae} as shown in Fig.~\ref{fig:core} (a)
and (b), respectively. Most of all, the nodes are much more densely
connected in {\it S. cerevisiae} than in {\it E. coli}. This
difference can be first ascribed to different mean connectivities of
the nodes in the core: it is about $1.47$ in {\it E. coli} and $2.65$
in {\it S. cerevisiae}.  However, a more striking difference exists in
their core organization. In {\it E. coli}, all $54$ circuits are
identified, all of which are single-element circuits representing
self-regulation. There are no circuits whose length (i.e the number of
edges on the cycle) is larger than $1$~\cite{thieffry98}. On the
contrary, only one or two single-element circuits are formed in its
randomized graphs. This organization of circuits in {\it
E. coli} is also contrasted with the one in {\it S. cerevisiae}. We
computed the shortest circuit for each pair of nodes in the core and
counted the numbers of node pairs for each given shortest-circuit
length.  The distribution of shortest-circuit length obtained for {\it
S. cerevisiae} is broad as shown in Fig.~\ref{fig:core} (c).  We
propose that the presence of single-element circuits in {\it
E. coli} is the main reason for the enhancement of $N_{\rm m}$ of {\it
E. coli} compared with both of its randomized graphs.  Once a node $i$
regulating itself is mutated, the input configurations to the
regulating rule $f_i$ are necessarily different between the
normal-mutant pair $(\Sigma,\hat{\Sigma})$ since it is guaranteed that
at least one of its regulators, the node $i$ itself, is
mutated. Recalling that a node can be mutated at the next time step
only if the input configurations from the normal-mutant pair are
different, one can see that single-element circuits have a higher
probability to be mutated than nodes which do not regulate themselves
[See Fig.~\ref{fig:tree-circuit} (c)]. Therefore networks with more
single-element circuits can be more adaptive.

In the core of {\it E. coli} network, $54$ edges are used for
single-element circuits and the remaining $30$ edges connect pairs of
distinct nodes. As a result, the network has many isolated nodes and
few small connected components, resulting in the rapid decay of the
relaxation time. In Fig.~\ref{fig:NmP} (c), we find that the
relaxation times observed in {\it E. coli} are mostly $1$ or $2$. From
this, we can analytically predict the value of $N_m$ as a function of
$\lambda$. Suppose $N_{\rm m}(t)$ saturates no later than time
$2$. From Eq.~(\ref{eq:sc}), $\bar{H}(1)=\lambda K N^{-1} + {\cal
O}(N^{-2})$ since $\bar{H}(0)=N^{-1}$ and
\begin{equation}
N_{\rm m}\simeq N H(2) \simeq N \lambda K \bar{H}(1)\simeq  
\lambda^2 K^2.
\label{eq:ecoliapprox}
\end{equation}
This is in good agreement with the true value as shown in
Fig.~\ref{fig:NmP} (a).

\begin{figure}
\includegraphics[width=\columnwidth]{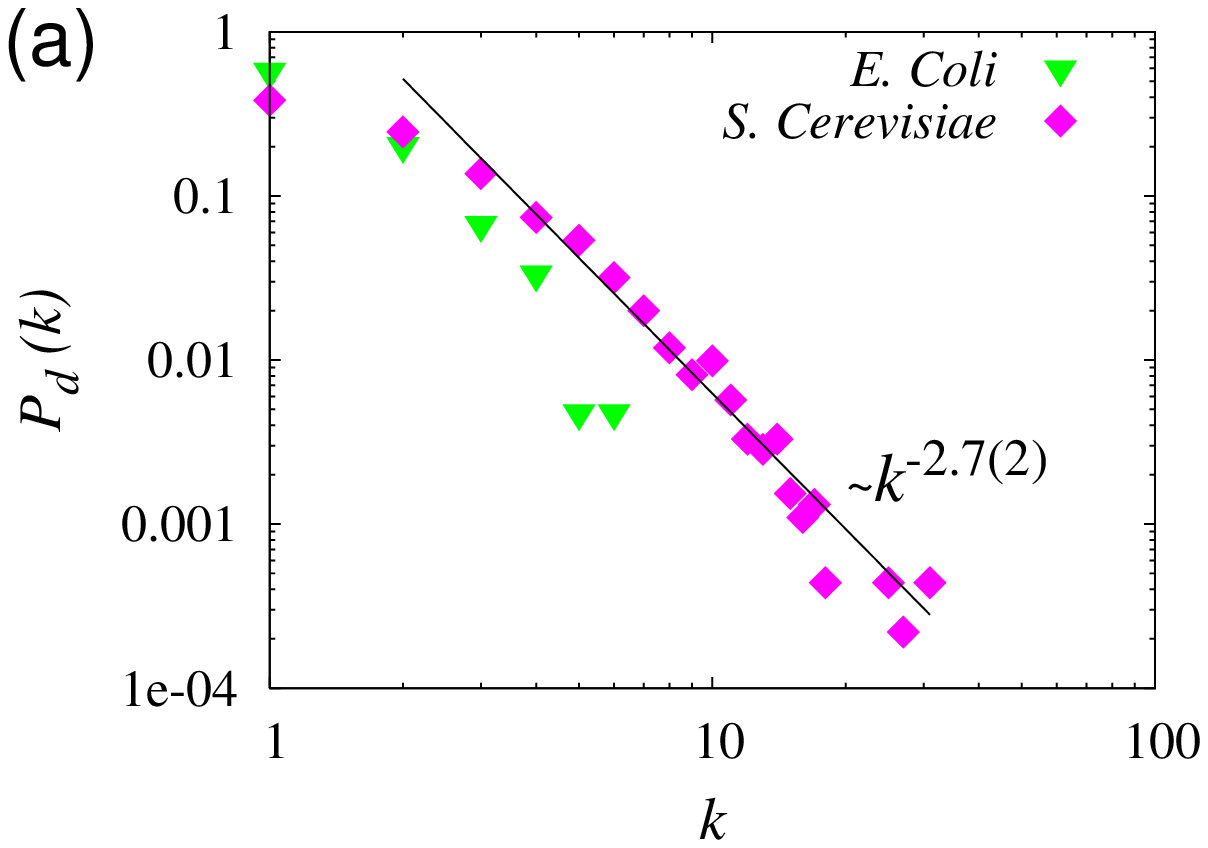}
\includegraphics[width=\columnwidth]{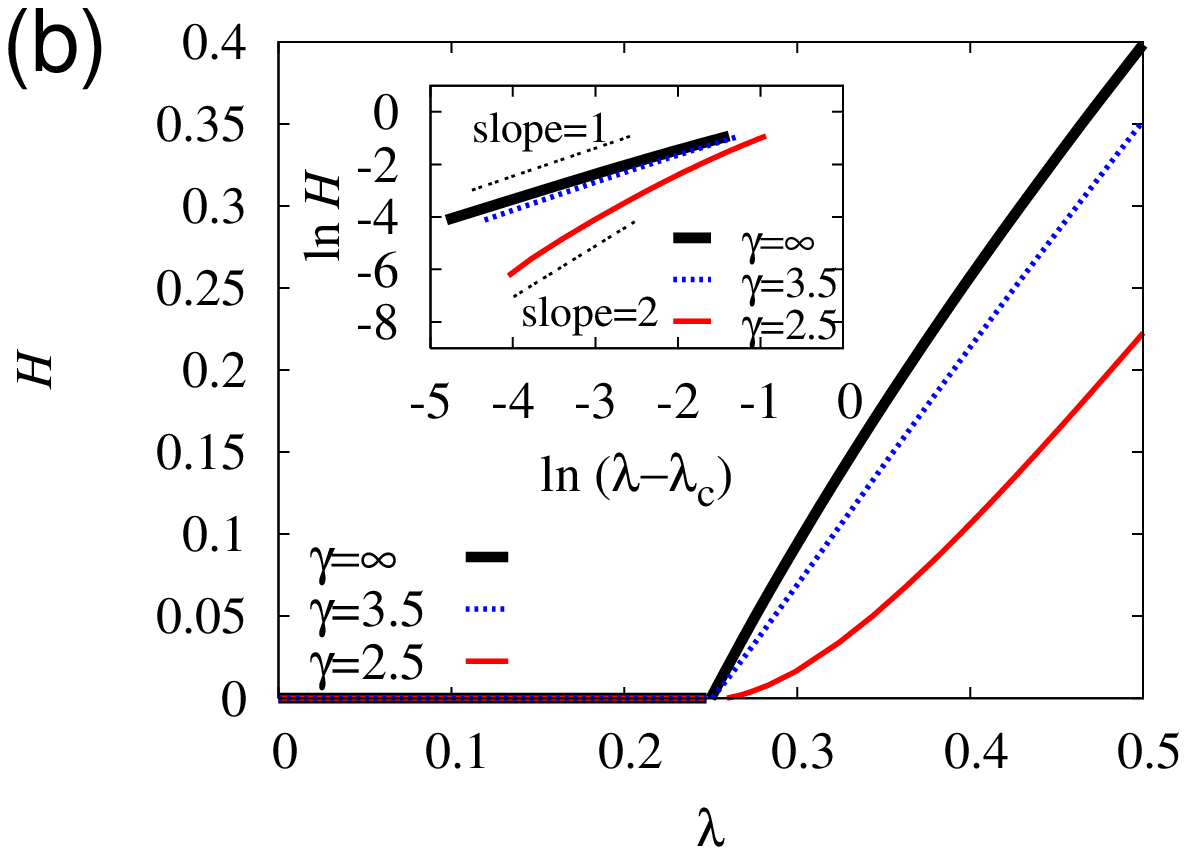}
\caption{Connectivity pattern and its effect on the critical behavior of the 
Hamming distance. (a) In-degree distributions $P_d(k)$ for {\it E. coli} and {\it S. cerevisiae}. 
For {\it S. cerevisiae}, its asymptotic behavior is a power-law, $P_d(k)\sim k^{-\gamma}$ 
with $\gamma\simeq 2.7(2)$. On the other hand, the observed values of $k$ are only up to $6$ 
and so it is hard to discern the functional form of $P_d(k)$ in {\it E. coli}. (b)  
Hamming distance $H$ as a function of $\lambda$ numerically obtained from 
Eq.~(\ref{eq:sc_simple}) with $P_d(k)$ of the static model~\cite{lee04}, which 
has a power-law tail as $P_d(k)\sim k^{-\gamma}$ with the exponent $\gamma$ tunable. 
The inset shows that 
$H\sim \Delta$ commonly for  $\gamma\to\infty$ and $\gamma=3.5$, and that $H\sim \Delta^2$ 
for $\gamma=2.5$, in agreement with Eq.~(\ref{eq:beta}).} 
\label{fig:critical}
\end{figure}

\subsection{Power-law in-degree distribution in {\it S. cerevisiae}}
In {\it S. cerevisiae}, the most significant dynamical feature that we
found and that we need to explain is the slower increase of $N_{\rm
m}$ with $\lambda$ as compared with the type-I randomized graph, shown
in Fig.~\ref{fig:NmP} (b). Contrary to the type-II randomized graphs,
those of type-I do not preserve the degree distribution of the
original network. From this, we can conjecture that the degree
distribution of {\it S. cerevisiae} causes the slow increase of
$N_{\rm m}$. To check this, we analyze in detail the dependence of the
Hamming distance on the degree distributions.

With uncorrelated in- and out-degree as is the case in the regulatory networks 
considered here, Eq.~(\ref{eq:sc}) is reduced to $H(t)=\bar{H}(t)$ and 
\begin{equation}
H(t+1) = \lambda \sum_k [1-(1-H(t))^k] P_d(k).
\label{eq:sc_simple}
\end{equation}
Thus the in-degree distribution $P_d(k)$ determines the behavior of
the Hamming distance $H(t)$. The in-degree distributions of {\it
E. coli} and {\it S. cerevisiae} shown in Fig.~\ref{fig:critical} (a)
are quite different from each other.  The maximum degree is $31$ in
{\it S. cerevisiae} while it is only $6$ in {\it E. coli}.
Furthermore, the log-log plot of $P_d(k)$ in {\it S. cerevisiae}
indicates that $P_d(k)\sim k^{-\gamma}$ with $\gamma\simeq
2.7(2)$. The functional form of $P_d(k)$ for {\it E. coli} is hard to
determine because of the small range for observable $k$ values.  Note
that the in-degree distribution of the type-I randomized graphs obey a
Poisson distribution, $P_d(k)=\langle k\rangle^k e^{-\langle
k\rangle}/k!$. Let us consider an in-degree distribution which has a
power-law tail, i.e., $P_d(k)\sim k^{-\gamma}$. Then, we find from
Eq.~(\ref{eq:sc_simple}) that the Hamming distance in the stationary
state behaves as $H\sim \Delta^\beta$ for $\lambda$ larger than the
critical value $\lambda_c$ with $\Delta\equiv \lambda/\lambda_c-1$ and
the critical exponent $\beta$ given by
\begin{equation}
\beta = \left \{
\begin{array}{ll}
1 & (\gamma>3),\\
1/(\gamma-2) & (2<\gamma<3). 
\end{array}
\right.
\label{eq:beta}
\end{equation}
The derivation of Eq.~(\ref{eq:beta}) is given in Appendix. 
We restricted the range of $\gamma$ to $\gamma>2$ because the mean 
connectivity diverges with $\gamma<2$. When the in-degree is subject 
to a Poisson distribution or an exponentially-decaying distribution, it 
corresponds to $\gamma\to\infty$ and the critical behavior is 
the same as that for $\gamma>3$. We present the numerical solution to 
Eq.~(\ref{eq:sc_simple}) in Fig.~\ref{fig:critical} (b) for 
$\gamma\to\infty$ (Poisson distribution), $\gamma=3.5$, and $\gamma=2.5$. 

The increase of $\beta$ with decreasing $\gamma$ below $\gamma=3$
indicates a difference in the behavior of the Hamming distance near
the critical point between networks with $\gamma>3$ and those with
$2<\gamma<3$. Suppose we have two networks with a power-law in-degree
distribution $P_d(k)\sim k^{-\gamma}$: One has $\gamma=3.5$ and the
other has $\gamma=2.5$, and both have $\langle k\rangle=4$. Then, in
the region $0<\Delta =\lambda/\lambda_c-1\ll 1$, the Hamming distance
behaves as $H\sim \Delta$ for $\gamma=3.5$ and $H\sim \Delta^2$ for
$\gamma=2.5$: the former increases more rapidly than the latter in the
region $\Delta\ll 1$. Also the region where the Hamming distance
remains non-zero but small, e.g., $H\leq 0.05$ is larger with
$\gamma=2.5$ than with $\gamma=3.5$: it is given by $\lambda\in
(0.25:0.29]$ with $\gamma=3.5$ and $\lambda\in (0.25:0.35]$ with
$\gamma=2.5$.  Such dependence of the Hamming distance on the
in-degree exponent $\gamma$ can thus explain different network
responses between {\it S. cerevisiae} and its type-I randomized
graphs. It is the broad in-degree distribution with $\gamma=2.7(2)$
that makes the number of mutated elements increase with $\lambda$ more
slowly than in the corresponding type-I randomized graphs that have
$\gamma\to\infty$. Due to such a slow increase of the Hamming
distance, {\it S. cerevisiae} can keep the size of mutation small for
a wider range of the parameter $p$ or $\lambda$, which would be much
larger with random structures.

\section{Conclusion}
We performed numerical experiments - spread of mutation
- to probe the dynamic stability of the recently-unveiled networks of
gene transcriptional regulation of {\it E. coli} and {\it
S. cerevisiae} and provided analytical confirmation for the results by
analyzing their structural features. While the small number of edges
per node in {\it E. coli} fundamentally prohibits a global spread of
mutation, a relatively large number of edges in {\it S. cerevisiae}
enables a global mutation conditionally depending on the regulating
rules. We further identified the relevant structural features which
are distinguished from those of random graphs: All circuits of the
regulatory network of {\it E. coli} are single-element circuits and
the in-degree distribution of {\it S. cerevisiae} takes a power-law
form. Single-element circuits in {\it E. coli} have higher probability
to be mutated than nodes without self-regulation. The broad in-degree
distribution in {\it S. cerevisiae} smoothens the increase of the
number of mutated elements. This increase would be sharper for an
exponential distribution, as is the case in the random graphs.

These biological networks appear to follow design principles that tend
to balance the size of mutation. The small mean connectivity of the
regulatory network of {\it E. coli} would restrict the size of
mutations drastically, which is compensated by the abundance of
single-element circuits that lead to the required enhancement of the
mutation size. In the case of {\it S. cerevisiae}, its global
characteristics of the regulatory network, a mean connectivity larger
than 2, would lead to a very large mutation size, but a very
heterogeneous interconnectivity pattern suppresses it. These local
structural features demonstrate that both genetic networks have
evolved, in spite of the restrictions imposed by the global
characteristics, in such a direction that they can stay dynamically
between stable (i.e., rarely mutated on a global scale) and unstable
(easily mutated).  Being neither stable nor unstable appears to be
necessary for living organisms to maintain their stable internal state
and adapt itself to fluctuating external environment
simultaneously. Therefore our finding suggests that such a marginal
dynamic stability of the whole system is supported by a selected
structural organization of the internal systems on smaller scales, as
the transcriptional regulatory network studied in this work. While we
have concentrated only on the average in-degree, the organization of
circuits, and the in-degree distribution of the network, further
structural analysis will be helpful to illuminate how structure
supports function.

\acknowledgements
We thank Uri Alon and Richard A. Young for allowing us to use their data.
This work was supported by Deutsche Forschungsgemeinschaft (DFG).

\appendix

\section{Derivation of Eq.~(\ref{eq:beta}) from Eq.~(\ref{eq:sc_simple})}

To find the behavior of $H=\lim_{t\to\infty} H(t)$ as a function of 
$\lambda$ near the critical point $\lambda_c=\langle k\rangle^{-1}$, 
we set $H(t+1)=H(t)=H$ and expand Eq.~(\ref{eq:sc_simple}) 
for small $H$, which leads to 
\begin{equation}
H=\lambda \sum_{n=1}^\infty \frac{(-1)^{n+1}\langle k^n\rangle}{n!} H^n.
\label{eq:expand}
\end{equation}
Here $\langle k^n\rangle$ is the $n$th moment of the in-degree 
distribution $P_d(k)$, i.e., $\langle k^n\rangle\equiv\sum_k k^nP_d(k)$. 
It is finite for all $n$ only if $P_d(k)$ decays exponentially. 
In this case, all the terms in the right-hand-side of Eq.~(\ref{eq:expand}) 
are analytic and keeping the first two leading terms, one finds 
that Eq.~(\ref{eq:expand}) is expressed as 
$H\simeq \lambda \langle k\rangle H - \lambda\langle k^2\rangle H^2/2$. 
This allows us to see that $H=0$ for $\lambda<\lambda_c=\langle k\rangle^{-1}$ 
and $H\sim \Delta$ with $\Delta \equiv (\lambda-\lambda_c)/\lambda_c$ 
for $\lambda>\lambda_c$. 

When the in-degree distribution is a power-law asymptotically, 
$P_d(k)\sim k^{-\gamma}$, all the moments $\langle k^n\rangle$ are 
not finite: $\langle k^n\rangle$ for $n>n_*$ with 
$n_*= \lceil\gamma-2\rceil$ diverges as 
$k_{\rm max}^{n-\gamma+1}/(n-\gamma+1)$, where $\lceil x\rceil$ 
is the smallest integer not smaller than $x$ and $k_{\rm max}$ is 
the (average) largest in-degree. The largest in-degree diverges 
as $N^{1/(\gamma-1)}$, which is derived from the relation 
$\sum_{k>k_{\rm max}} P_d(k) \sim N^{-1}$. Thus 
$\langle k^n\rangle \sim N^{(n-\gamma+1)/(\gamma-1)}$. 
Such diverging terms are arranged as  
$H^{\gamma-1} \sum_{n>n_*} (-1)^{n+1} [k_{\rm max} H]^{n-\gamma+1}/
[n!(n-\gamma+1)]$ in the right-hand-side of Eq.~(\ref{eq:expand}). 
Here the summation converges to a constant in the limit 
$k_{\rm max}\bar{H}\to\infty$ due to alternating signs and 
fast decay of the coefficients~\cite{lee05}. Thus the small-$H$ 
expansion  of Eq.~(\ref{eq:expand}) reads as 
$H = \lambda \sum_{n=1}^{n_*} (-1)^{n+1} \langle k^n\rangle H^n/n! 
+ \lambda ({\rm constant}) H^{\gamma-1} + \cdots.$.  
The $H^{\gamma-1}$ term is relevant to the critical behavior of $H$ 
for  $\gamma<3$ since it holds for $\gamma<3$ that 
$H\simeq \lambda \langle k\rangle H + \lambda ({\rm const.}) H^{\gamma-1}$, 
yielding $H\sim \Delta^{1/(\gamma-2)}$. On the other hand, the linear 
and quadratic terms are relevant for $\gamma>3$ as for exponentially-decaying 
in-degree distributions. In summary, the Hamming distance $H$ 
with a power-law in-degree distribution $P_d(k)\sim k^{-\gamma}$ 
behaves near the critical point as 
\begin{equation}
H \sim \left\{
\begin{array}{cc}
\Delta & (\gamma>3),\\
\Delta^{1/(\gamma-2)} & (2<\gamma<3).
\end{array}
\right.
\label{eq:critical}
\end{equation}

\end{document}